\documentclass[a4paper, 11pt]{article}
\pdfoutput=1

\usepackage{amsfonts}
\usepackage{amsmath}
\usepackage{amssymb}
\usepackage{subcaption}
\usepackage{color}
\usepackage{graphicx}
\usepackage{authblk}
\usepackage[authoryear]{natbib}
\usepackage{booktabs}
\usepackage[top=1in, bottom=1.25in, left=60pt, right=60pt]{geometry}

\newcommand{\wvec}{\left[\hspace{-4pt}\begin{array}{c} \mathbf{a}_1^\intercal \\[2pt] \mathbf{a}_2^\intercal	\end{array}\hspace{-4pt}\right]} 
\newcommand{\wvectwo}{\left[\hspace{-4pt}\begin{array}{c} \mathbf{a}_2^\intercal \\[2pt] \mathbf{a}_1^\intercal	\end{array}\hspace{-4pt}\right]} 

\newcommand{\revone}[1]{{ #1}}
\newcommand{\revtwo}[1]{{ #1}}

\begin{document}
\title{\revtwo{EEG in the classroom:} Synchronised neural recordings during video presentation}
\author[1,+,*]{Andreas Trier Poulsen}
\author[1,+]{Simon Kamronn}
\author[2,3]{Jacek Dmochowski}
\author[3]{Lucas C. Parra}
\author[1]{Lars Kai Hansen}

\affil[1]{Technical University of Denmark, DTU Compute, Kgs.\ Lyngby, Denmark}
\affil[2]{Stanford University, Department of Psychology,  Palo Alto, USA}
\affil[3]{City College of New York, Department of Biomedical Engineering, New York, USA}
\affil[*]{atpo@dtu.dk}
\affil[+]{these authors contributed equally to this work}

\maketitle

\begin{abstract}
We performed simultaneous recordings of  electroencephalography (EEG) from multiple students in a classroom, and measured the inter-subject correlation (ISC) of activity evoked by a common video stimulus.
\revtwo{The neural reliability, as quantified by ISC, has been linked to engagement and attentional modulation in earlier studies that used high-grade equipment in laboratory settings. Here we reproduce many of the results from these studies using portable low-cost equipment, focusing on the robustness of using ISC for subjects experiencing naturalistic stimuli.}
\revtwo{The present data shows that stimulus-evoked neural responses, known to be
modulated by attention, can be tracked in for groups of students with synchronized EEG acquisition. This is a step towards real-time inference of engagement in the classroom.}
\end{abstract}

\flushbottom

\thispagestyle{empty}

\section*{Introduction}
Engagement and attention are important in situations of learning, but most methods for measuring of attention or engagement are intrusive and  unrealistic in everyday situations \citep{robinson1997,cohen1990,radwan2005}.
Recently, inter-subject correlation (ISC) of electroencephalography (EEG) has been proposed as a marker of attentional engagement  \citep{Dmochowski2012,Dmochowski2014,ki2016attention} and we ask in this work whether it can be recorded robustly with commercial-grade wireless EEG devices in a classroom setting. Furthermore, we address two other issues related to the robustness of the signal: The potential neurophysiological origin of the measure and the robustness of the detection scheme to inter-subject variability in spatial alignment.

User engagement has been defined as `... the emotional, cognitive and behavioural connection that exists, at any point in time and possibly over time, between a user and a resource'  \citep{attfield2011towards}. Traditional approaches to measuring engagement  are based on capturing user behaviour via user interfaces, self-report, or manual annotation  \citep{o2013examining}. However, tools from cognitive neuroscience are increasingly being employed  \citep{szafir2013artful}.
Recent efforts in neuroscience aim to elucidate perceptual and cognitive processes in a more realistic setting and using naturalistic stimuli  \citep{Dmochowski2012,ringach2002receptive,hasson2004intersubject,Lahnakoski2014,lankinen2014,chang2015}.
From an educational perspective such quantitative measures may help identify mechanisms that make learning more efficient  \citep{szafir2013artful}, align services better with students needs  \citep{attfield2011towards}, or monitor critical task performance  \citep{lin2013can}. The potential uses of engagement detection in the classroom are numerous, e.g., real-time and summary feedback for the teacher, motivational strategies for increased student engagement, and screening for impact of teaching materials.
Before the findings of tracking attentional responses with neural activity \citep{Dmochowski2012,Dmochowski2014,ki2016attention} can be employed in a real-time classroom scenario, several issues must be addressed first, including: 1) Is it possible to reproduce the ISCs to naturalistic stimuli under the adverse conditions of a classroom? 2) Are the ISCs robust to inter-student variability of the spatial information processing networks? And 3) can ISCs be recorded with equipment that is both comfortable and affordable enough to make it a realistic technology for schools?

Here we investigate the feasibility of recording such neural responses from students who are viewing videos. We use an approach developed by Dmochowski et al.\ (2012) that uses inter-subject correlation (ISC) of EEG evoked responses.
The basic premise is that subjects who are engaged with the content exhibit reliable neural responses that are correlated across subjects and repetitions within the same subject.  In contrast, a lack of engagement manifests in generally unreliable neural responses  \citep{ki2016attention}.
ISC of neural activity while watching films have been shown to predict the popularity and viewership of TV-series and commercials \citep{Dmochowski2014},
and shows clinical promises as a measure of consciousness levels in non-responsive patients \citep{Naci2015} (fMRI study). We argue here that the neural reliability of students indeed may be quantified on a second-by-second basis in groups and in a classroom setting, and we seek to investigate the robustness of measuring it with electroencephalography (EEG) responses during exposure to media stimuli.

\begin{figure}
\centering
\includegraphics[width=\columnwidth]{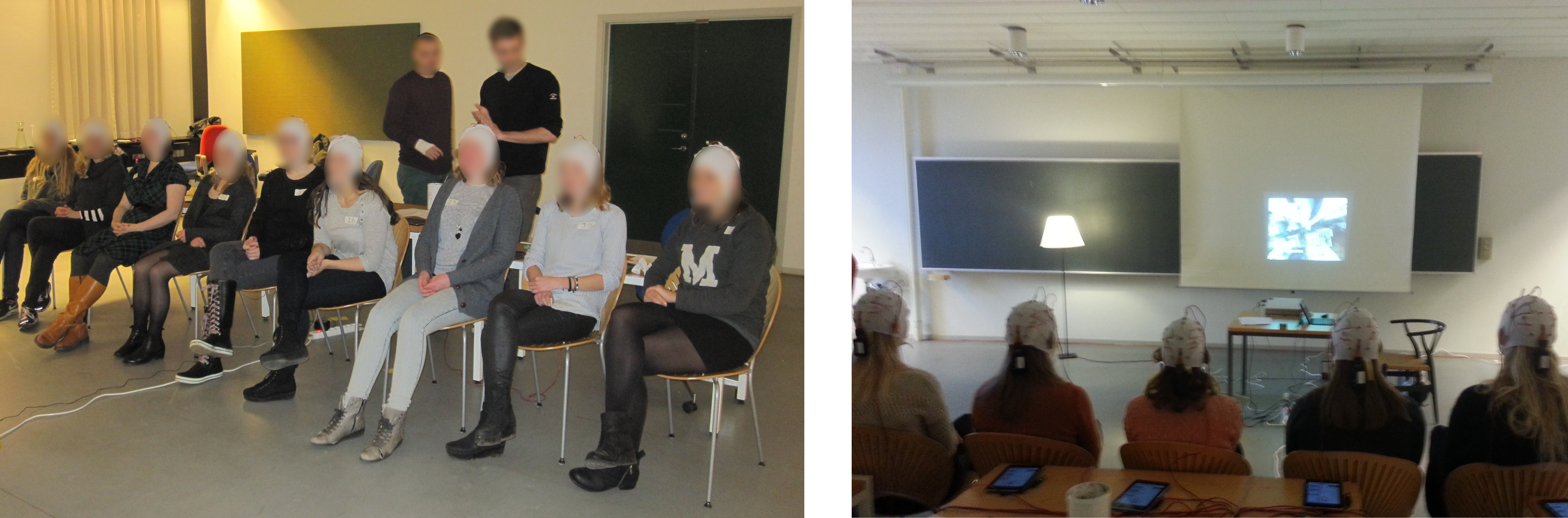}%
\caption{\revone{Experimental setup for joint viewings. \textbf{(Left):} 9 subjects where placed on a line to induce a cinema-like experiences. \textbf{(Right):} Subjects seen from the back, watching films projected onto a screen.
Tablets recording EEG are resting on the tables behind the subjects. The signal is  transmitted wirelessly from each subject.}}
\label{Fig.setup_joint}
\end{figure}

To enable correlations between multi-dimensional EEG, correlated component analysis (CorrCA) was introduced \citep{Dmochowski2012}. CorrCA finds multiple spatial projections that are shared amongst subjects, such that their components are maximally correlated across time.
Here we are interested in the reproducibility of using CorrCA as a measure of inter-subject correlation, and  will focus predominantly
on the first component, which captures most of the neural responses shared across students.

The main goal of the present work is to determine \revtwo{whether student neural reliability} can be quantified in a real-time manner \revtwo{based on} recordings of brain activity in a classroom setting using a low-cost, portable EEG system -- the Smartphone Brain Scanner
 \citep{stopczynski2014smartphone}. \revtwo{With regard to} the robustness of the detection scheme, we report on both theoretical and experimental investigations.
\revone{First, we show that ISC evoked by rich naturalistic stimuli is robust enough to be reproduced with commercial-grade equipment, and to be recorded simultaneously from multiple subjects in a classroom setting. \revtwo{This opens up for} the possibility of real-time estimation of student \revtwo{attentional} engagement.
Secondly, we show mathematically that the CorrCA algorithm is surprisingly robust to variations in the spatial patterns of brain activity across subjects.
Finally, we demonstrate that the level of ISC
is related to a very basic visual response that is modulated by
narrative coherence of the video stimulus.}

\section*{Results}
To monitor \revtwo{neural reliability} we used video stimuli as they provide a balance between realism and reproducibility  \citep{hasson2004intersubject}. We recorded EEG activity using the
Smartphone Brain Scanner while subjects watched short video clips of approximately 6 minutes duration, either individually or in a group setting (Fig. \ref{Fig.setup_joint}). 
To measure reliability of EEG responses, we used correlated components analysis (CorrCA, see Methods) to extract maximally correlated time series with shared spatial projection across repeated views within the same subject (inter-viewing correlation, IVC), or between subjects (inter-subject correlation, ISC).

One of our main points of interest is to investigate the robustness of ISC from EEG recorded in a classroom through comparisons with results previously measured in a laboratory setting \citep{Dmochowski2012}. We therefore employed similar methods of analysis and calculated ISCs and IVCs in 5 second windows with 80 \% overlap to investigate their temporal development in a 1-second resolution. We chose to analyse the EEG with CorrCA in a broad frequency band (0.5 and 45 Hz), instead of investigating specific frequency bands, to keep the analysis methods comparable with the prior lab-based study.
Moreover, CorrCA is a method used for robustly measuring ISC with low computational costs; hence making it a good candidate for long term real-time analyses on small devices in a classroom setting.

\begin{figure}
\centering
\includegraphics[width=\columnwidth]{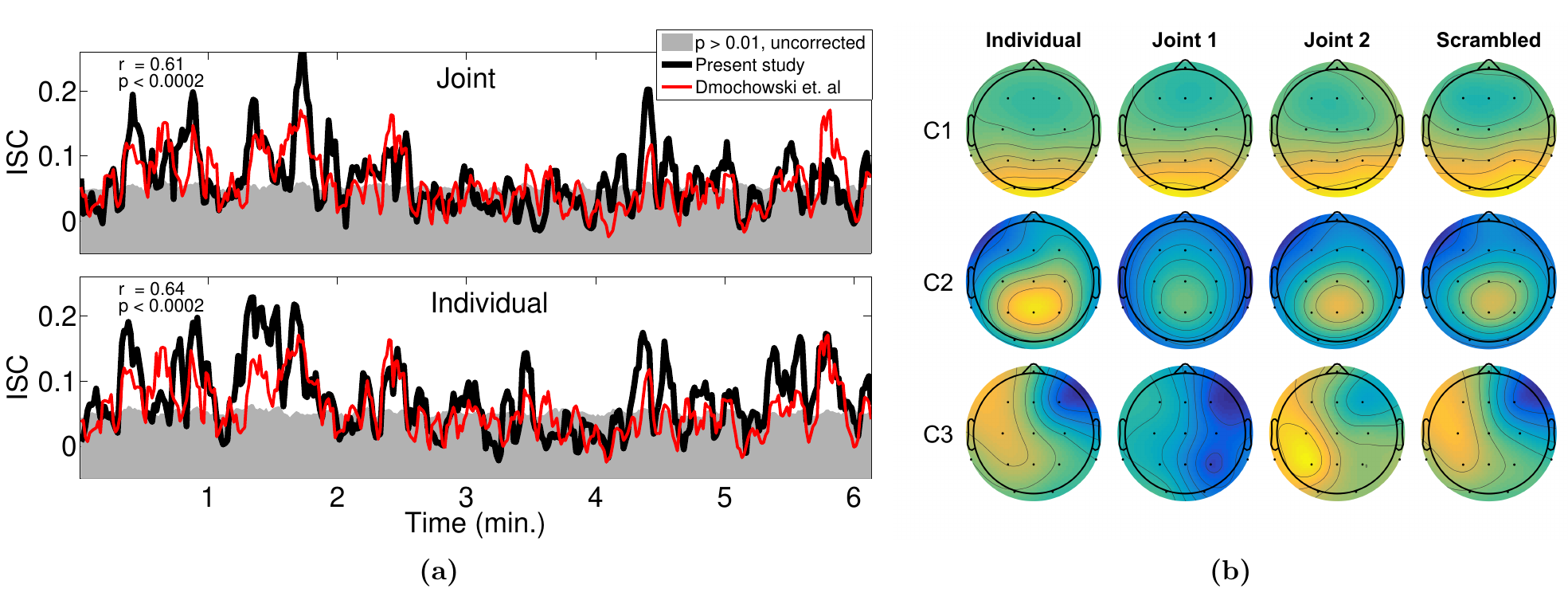}%
\caption{\revone{ISC of neural responses to naturalistic stimuli are robust across different groups of subjects and reproducible in a classroom setting.} \textbf{(a)} Comparison between the ISC obtained by Dmochowski et al. 2012 and the present study \revone{for the first CorrCA component and the first viewing of \textit{Bang! You're Dead}. The ISC is calculated with a 1-second resolution (5 s windows, 80\% overlap).
The grey area indicates chance levels for ISC ($p > 0.01$ estimated with time-shuffled surrogate data, uncorrected for multiple comparisons). } \textbf{(b)} \revone{The corresponding scalp projections of the first three components obtained from the correlated component analysis (CorrCA) of each of the four subject groups watching \textit{Bang! You're Dead} the first time. For each component, CorrCA finds one shared set of weights for all subjects in the group.} Four distinct groups of subjects watched videos in different scenarios: individually on a tablet computer (\textit{Individual}), individually with order of scenes scrambled in time (\textit{Scrambled}), \revone{and jointly in a classroom as seen in Fig. \ref{Fig.setup_joint} (\textit{Joint 1} and \textit{Joint 2}).} For each projection, the polarity was normalized so the value at the Cz electrode is positive.}
\label{Fig.iscscalp}
\end{figure}

\revtwo{The subjects watched three video clips, which were presented twice in random order. The first video was}
a suspenseful excerpt from the short film,  \textit{Bang! You're Dead}, directed by Alfred Hitchcock. It was selected because it is known to effectively synchronize brain responses across viewers \citep{hasson2008neurocinematics,Dmochowski2012}.
The second video was an excerpt from \textit{Sophie's Choice}, directed by Alan J. Pakula (1982), and the third was an uneventful baseline video of people silently descending an escalator.
For both the joint and individual recording scenarios, the time course of the ISC, based on the first CorrCA component from subjects watching the film, closely reproduces results obtained previously in a laboratory setting (Fig. \ref{Fig.iscscalp}a and Table \ref{Tab.isccorr_both}).

\begin{figure}
\centering
\includegraphics[width=.85\columnwidth]{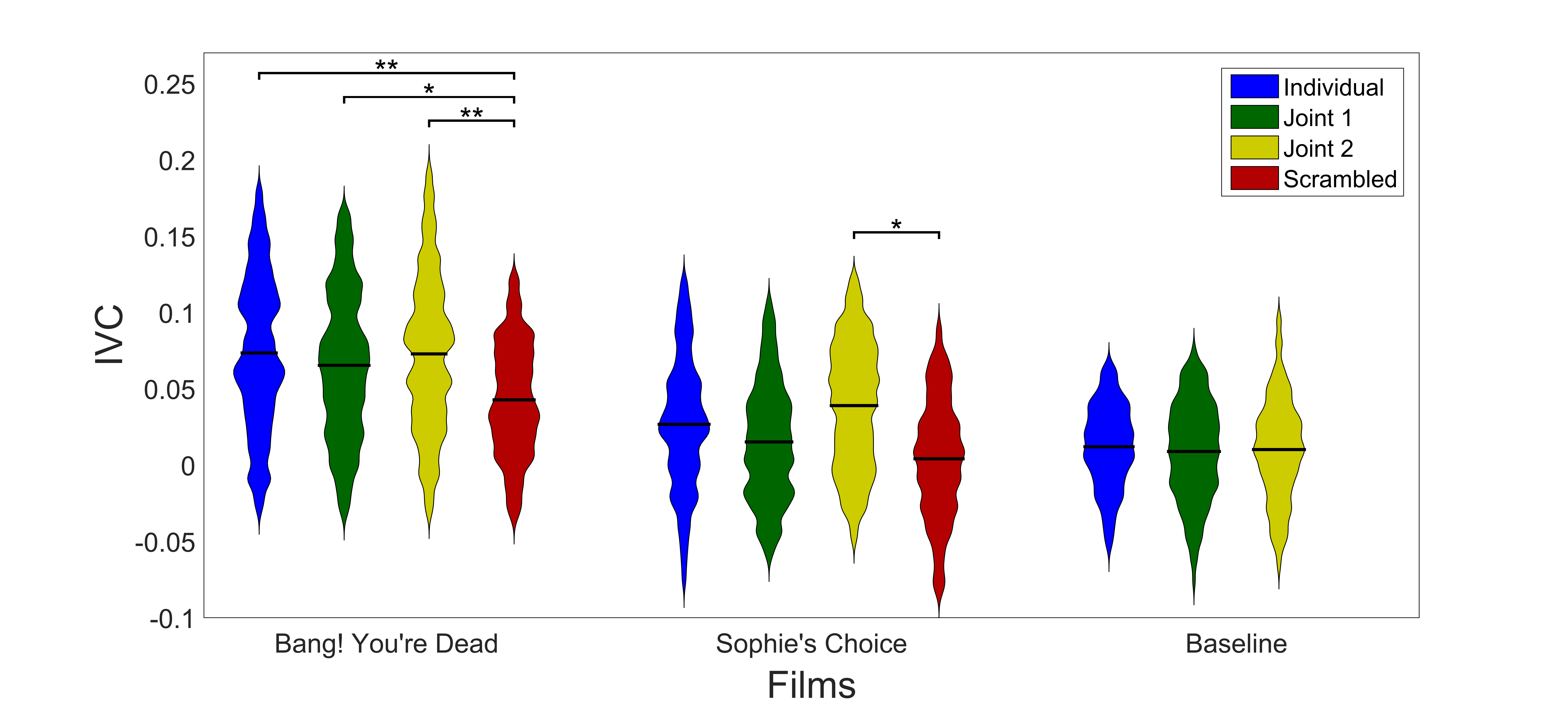}%
\caption{\revone{Distribution and mean of IVC calculated from the first CorrCA component for subject groups and films. Violin plots show distributions of IVC estimated using a squared exponential (normal) kernel with bandwidth of 0.005 \citep{hoffmann2015}. Horizontal black bars denote distribution means. For  visualisation purposes, the extreme 2.5\% values at either end of the distributions were left out of the violin plots (but were kept for estimating mean and p-values).}
A block permutation test (block size $B = 25\ s$) \revtwo{was employed} to estimate statistical significant differences \revone{in the mean IVC} between viewing conditions \revone{(uncorrected for multiple comparisons).} \revone{For both films there were significant differences in mean IVC between groups with normal narrative and the \textit{Scrambled} group (\textit{Bang! You're Dead}: $p_{\textit{Individual}} = 0.006$, $p_{\textit{Joint 1}} = 0.033$, $p_{\textit{Joint 2}} = 0.004$; \textit{Sophie's Choice}: $p_{\textit{Individual}} = 0.059$, $p_{\textit{Joint 1}} = 0.37$, $p_{\textit{Joint 2}} = 0.012$). However, there were no significant differences between groups with the original, unscrambled narrative.}
 Note that the \textit{Scrambled} group did not watch the baseline video.}
\label{Fig.barintraplot}
\end{figure}

\revtwo{An indication of the stability of the technique is provided by the spatial patterns of the neural activity that drives these reproducible responses.
Similar to other component extraction techniques, such as independent component analysis or common spatial patterns \citep{parra2003,koles1990}, CorrCA reduces the signal of multiple electrodes to a few components.
The ISC is then computed for the first few components,} \revone{which capture most of the correlation between recordings.}
The strongest \revone{three} correlated components show a stable pattern of activity across the different groups and recording
\revone{conditions (Fig. \ref{Fig.iscscalp}b), all three obtaining significant \revtwo{spatial} correlations between groups ($r_{\textit{comp1}} = 0.97$, $r_{\textit{comp2}} = 0.91$, $r_{\textit{comp3}} = 0.79$, all with $p < 0.002$ for uncorrected permutation test), for \textit{Bang! You're Dead}.
The robustness to recording conditions is also apparent for the second film clip from \textit{Sophie's Choice} ($r_{\textit{comp1}} = 0.51$, $p < 0.002$; $r_{\textit{comp2}} = 0.48$, $p = 0.008$; $r_{\textit{comp3}} = 0.36$, $p = 0.033$), albeit with a lower average correlation, which for the first two components may be due to noisy scalp maps for the \textit{Joint 1} group and \textit{Individual} group, respectively (see supplementary Fig. S1). For the baseline video, only the first component achieved significant average correlation between groups ($r_{\textit{comp1}} = 0.46$, $p = 0.014$).}
\revone{The lower stability in the scalp maps obtained for \textit{Sophie's Choice} and the baseline video could be explained by the lower ALD of these stimuli (see below), since these films obtain lower average IVC compared to \textit{Bang! You're Dead} for all groups (Fig. \ref{Fig.barintraplot}).}

\begin{table}
\caption{Correlation coefficients between the \revone{ISC time courses} obtained in a laboratory setting  \citep{Dmochowski2012} and those obtained in the present study (groups \textit{Individual, Joint 1} and \textit{Joint 2}). Inter-subject correlation (ISC) measures similarity of responses between subjects for the first and second viewings (v1,v2), and the inter-viewing correlation (IVC) measures similarity within-subject between the two views. \revone{Coefficients are calculated for the first CorrCA component recorded while watching \textit{Bang! You're dead}. **: $p<0.01$.}}
\label{Tab.isccorr_both}
\centering
\begin{tabular}{lccc}
 & ISC v1 & ISC v2 & IVC \\
\toprule
Individual & 0.64** & 0.33** & 0.49** \\
\hline
Joint group 1 & 0.51** & 0.15** & 0.44** \\
\hline
Joint group 2 & 0.61** & 0.28** & 0.54** \\
\bottomrule
\end{tabular}
\end{table}

\begin{table}
{\small
\caption{\revone{\revtwo{Scenes described by the subjects as having the strongest impression on them.} Based on the 30 subjects which saw \textit{Bang! You're Dead} with uninterrupted narrative. \revtwo{In a post-experiment questionnaire, subjects were asked to describe the scenes that made the strongest impression on them. Their answers were collected} in the eight groups. The subjects each mentioned 1.77 scenes on average (0.77 std.). 29 subjects (97 \%)  mentioned either \revtwo{scenes where the boy points the gun} at his mother or at other people.}}
\label{Tab.questbang}
\centering
\begin{tabular}{llr}
\textbf{Scene} & \textbf{Approx.\ times} & \textbf{No of times mentioned (\%) }\\
\toprule
The boy shoots (or points gun at) mother & 2:25 and 3:00 & 16 (53 \%)  \\
\hline
The boy shoots (or points gun at) at people & 2:10, 3:30 and 5:30 &  15 (50 \%)  \\
\hline
The boy loads another bullet into gun & 6:10 & 8 (27 \%)  \\
\hline
The uncle discovers his gun is gone & 4:35 & 4 (13 \%)  \\
\hline
The boy finds and loads gun & 0:25 and 1:40 & 4 (13 \%)  \\
\hline
The boy points at mirror or shoot towards camera & 0:40, 1:50 and 5:25 & 4 (13 \%)  \\
\hline
When the father did not run after the boy & 3:00 & 1 (3 \%)  \\
\hline
The abrupt ending & 6:14 & 1 (3 \%)  \\
\bottomrule
\end{tabular}
}
\end{table}

\revone{\revtwo{Previous research has indicated the potentials of ISC as a marker of engagement of conscious processing \citep{Dmochowski2012,Dmochowski2014,Naci2015,Lahnakoski2014,ki2016attention}. To further investigate this, we asked subjects post-experiment to describe the film segments (or "scenes") that made the biggest impact on them. We quantified their answers by assigning each answer to one of eight general scene descriptions.} Table \ref{Tab.questbang} shows that the scenes most frequently mentioned are "Boy pointing gun at mother" or "Boy pointing gun at people", and 29 out of 30 subjects mentioned one or both of the scenes as \revtwo{having had high} impact on them. The most frequently mentioned scene occurs around 2:25, where a peak in the ISC can be seen (Fig. \ref{Fig.iscscalp}a). The high impact of this particular scene was  confirmed by the suspense ratings presented in Naci et al. (2015).} 
\revtwo{See Dmochowski et al. (2012) for additional descriptions and examples of scenes eliciting high ISC in \textit{Bang! You're Dead}.}

To determine if the portable equipment, which uses only 14 \revone{channels}, can detect varying levels of \revtwo{neural reliability}, a second group of subjects watched \revone{the same two film clips} individually, but now with scenes scrambled in time. \revone{This intervention is a widely used tool to create a baseline with similar low-level stimuli, yet reduced engagement  \citep{miller1950verbal,anderson2006cortical,hasson2008neurocinematics,Dmochowski2012}}.
\revone{See Methods for more information on the definition and time scales of the scrambled scenes.}
Despite using consumer-grade EEG we find that IVC is significantly above chance for a large fraction of the original engaging clip, but drops dramatically when the scenes are scrambled in time (mean IVC, Fig. \ref{Fig.barintraplot}, $p<0.01$, for \textit{Bang! You're Dead}). Also the baseline video, \revtwo{which subjects reported not to engage them at all, only obtained significant ISC ($p < 0.01$, uncorrected) in 2.3 \% of the 354 tested time windows, compared to the 54.1 \% significant windows obtained for \textit{Bang! You're Dead}.}

For experiments conducted in \revone{less controlled, everyday} settings as in \revtwo{this study}, it is important to assess across-session reproducibility. To test this, we recorded a second group of subjects in a classroom setting \revtwo{who watched} the material together (\textit{Joint 1} and \textit{2}). These two groups obtained mean IVCs comparable to the individual recordings (Fig. \ref{Fig.barintraplot}, \textit{Bang! You're Dead}: $p>0.49$, \textit{Sophie's Choice}: $p>0.26$), and also showed reproducibility between the groups of simultaneous recordings (Fig. \ref{Fig.barintraplot}, \textit{Bang! You're Dead}: $p>0.49$, \textit{Sophie's Choice}: $p>0.08$).

Robustness to inter-subject variations in the spatial brain structure is a basic question when applying CorrCA to classroom data.
\revtwo{CorrCA is derived under the assumption that the spatial networks of subjects are identical. This assumption} could be challenged by inter-individual differences, however, it turns out to be surprisingly robust to such variability \citep{kamronn2015}.
To demonstrate this, we briefly analyse a 'worst case' scenario in which the true mixing weights of two subjects form a pair of \textit{orthogonal} vectors. The observations are assumed to consist of a single true signal, \textbf{z}, mixed into $D$ dimensions with additive Gaussian noise; $\textbf{X}_1 = \textbf{a}_1\textbf{z}^\intercal + \boldsymbol{\epsilon}\ $, $\textbf{X}_2 = \textbf{a}_2\textbf{z}^\intercal + \boldsymbol{\epsilon}\ $. Given a large sample, the covariance matrices are given as $\textbf{R}_{11} = P\cdot\textbf{a}_1\textbf{a}_1^\intercal  + \sigma^2 \textbf{I}\ $, $\quad \textbf{R}_{12} = P\cdot\textbf{a}_1\textbf{a}_2^\intercal\ $, where $P$ is the variance of \textbf{z} and $\sigma^2$ signifies the noise variance. For simplicity the weight vectors are assumed to be unit length. The two matrices in Eq.\ (\ref{eq.correlated component analysis}) can then be written as

\begin{align}
(\textbf{R}_{11} + \textbf{R}_{22})^{-1} = \frac{1}{P}\left([\textbf{a}_1\ \textbf{a}_2]  \wvec  + \frac{2\sigma^2}{P} \textbf{I}\right)^{-1}; \ \ \ \ \textbf{R}_{12} + \textbf{R}_{21} = P\cdot [\textbf{a}_1\ \textbf{a}_2]  \wvectwo , \label{eq.r12r21}
\end{align}

using block matrix notation.  With $\textbf{a}_1^\intercal\textbf{a}_2 = 0$, $\|\textbf{a}_1\|^2 = \|\textbf{a}_2\|^2 = 1$ and the Woodbury identity, the product of the two matrices in Eq.\ (\ref{eq.r12r21}) can be expressed as

\begin{align}
\left(\textbf{R}_{11} + \textbf{R}_{22}\right)^{-1} \left(\textbf{R}_{12} + \textbf{R}_{21}\right) = \frac{P}{2\sigma^2 + P}(\textbf{a}_1\textbf{a}_2^\intercal + \textbf{a}_2\textbf{a}_1^\intercal ). \label{eq.cocamatrix}
\end{align}

An eigenvector of matrix (\ref{eq.cocamatrix}) takes the form $\alpha\textbf{a}_1 + \beta\textbf{a}_2$,
with $\alpha = \pm\beta$ and $\pm\frac{P}{2\sigma^2 + P}$ as eigenvalues.
\revtwo{By applying this eigenvector to observations, $\textbf{X}_1$ and $\textbf{X}_2$, we see that CorrCA still identifies the relevant time series, \textbf{z}.}

\begin{figure}
\centering
\includegraphics[width=.75\columnwidth]{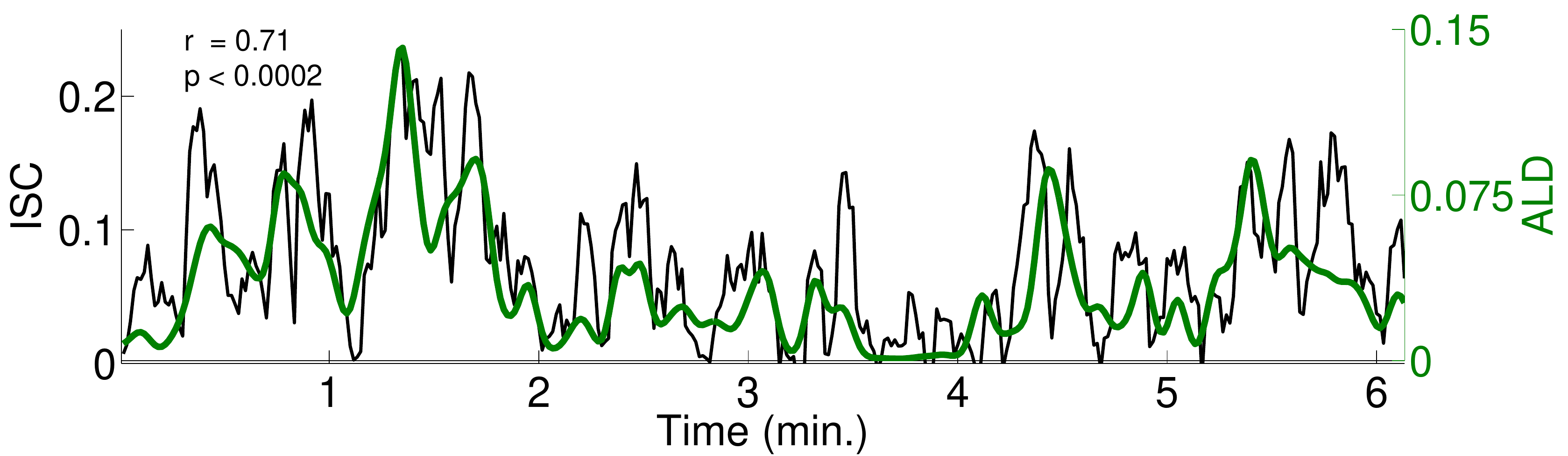}%
\caption{The ISC of \revone{the first CorrCA component is temporally correlated with the average luminance differences (ALD) of the film stimulus. ALD is calculated as the frame-to-frame difference in pixel intensity, smoothed to match the 5 s window of ISC, and mainly reflects the frequency of changes in camera position.} Data computed from the neural responses \revone{of subjects watching }\textit{Bang You're Dead}.}
\label{Fig.ISCscene}
\end{figure}

\begin{table}
\caption{Correlation coefficients between the ALD and the ISC for the
  two viewings (v1,v2) as well as the IVC for the first correlated
  component. The correlation is presented for \textit{Bang You're dead} and \textit{Sophie's Choice} for the \textit{Individual} and \textit{Scrambled (Scr)} groups. **: $p<0.01$.}
\label{Tab.scenecorr}
\centering
\begin{tabular}{lccc}
 & ISC v1 & ISC v2 & IVC \\
\toprule
Bang You're Dead & 0.71** & 0.61** & 0.56** \\
\hline
Sophie's Choice & 0.50** & 0.24** & 0.23** \\
\hline
Bang You're Dead (Scr) & 0.54** & 0.45** & 0.35** \\
\hline
Sophie's Choice (Scr) & 0.42** & 0.01 & -0.22** \\
\bottomrule
\end{tabular}
\end{table}

For the first CorrCA component, the channels weighted most heavily are the ones positioned over the occipital lobe (see Fig. \ref{Fig.iscscalp}b).
To estimate how much of the ISC was driven by basic low-level visual processing, we analysed the relation between ISC and a measure of frame-to-frame luminance fluctuations (average luminance difference, ALD; see methods). Note that to avoid synchronised eye artefacts and to ensure that only signals of neural origin contributed to the measured correlations, we removed independent components related to eye artefacts from the EEG (see methods).

\begin{figure}
\centering
\includegraphics[width=.95\columnwidth]{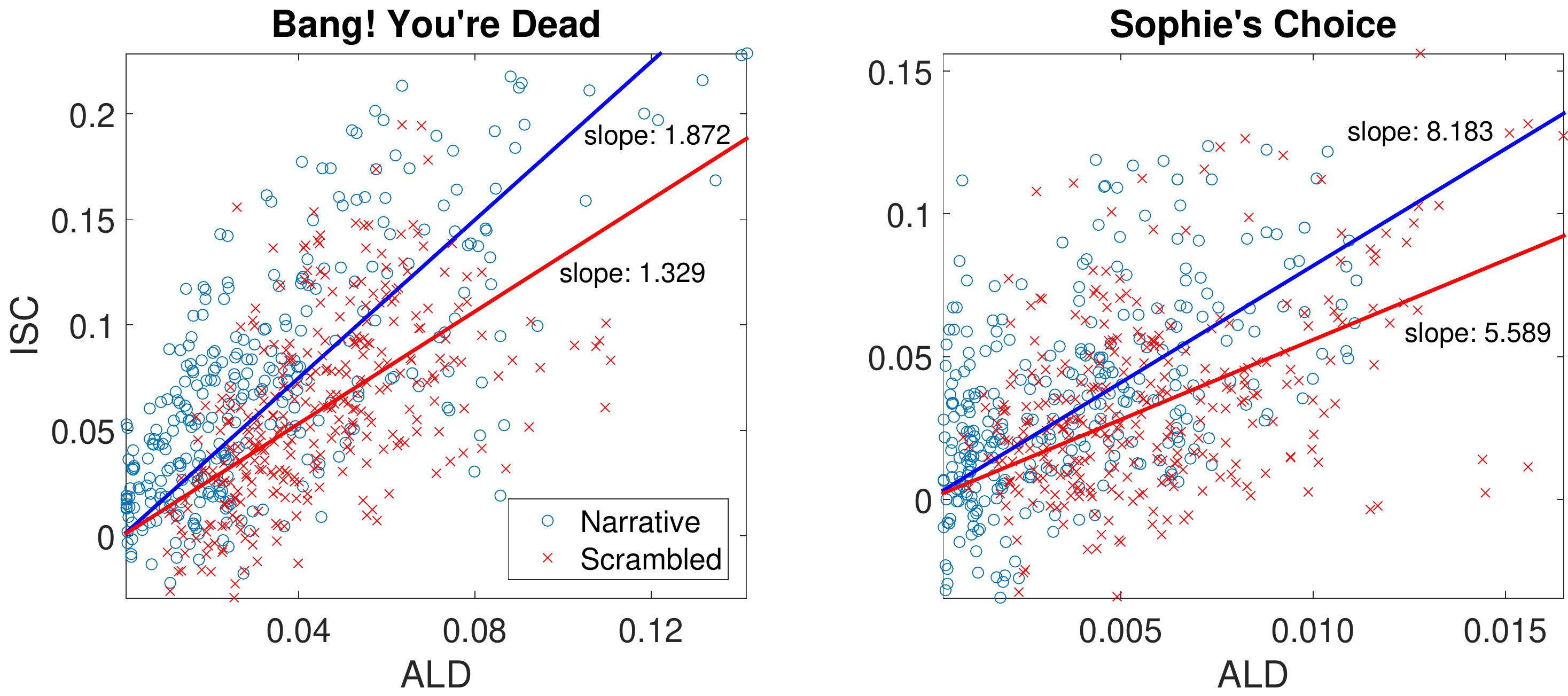}%
\caption{Relation between the ISC and the ALD for different conditions. Each point indicates
\revone{a point in the ISC time course as seen in Fig. \ref{Fig.iscscalp}a (5 s windows, 80\% overlap) and the corresponding} ALD calculated from the visual stimulus.
It is evident \revone{that time points} with higher luminance fluctuations (hight ALD) result in higher correlation of brain activity across subjects (high ISC).
\revone{The indicated "slope" is a least squares fit of the slope of lines passing through (0,0).}
The slope indicates the strength of ISC for a given ALD value. \revone{For both films} there is a significant drop in the slope ($p<0.01$: block permutation test with block size $B=25{\rm sec}$), thus the original narrative (blue) elicits higher ISC than the less engaging scrambled version of the films (red). \revone{Note that brightness of the scenes in \textit{Sophie's Choice} is much lower than in \textit{Bang! You're dead}, resulting in an ALD that is lower by almost a factor 10.}}
\label{Fig.GAINS}
\end{figure}

Figure \ref{Fig.ISCscene} and Table \ref{Tab.scenecorr} show that there is a significant correlation between the ISC and the ALD for both \textit{Bang! You're Dead} and \textit{Sophie's Choice} \revone{for the first CorrCA component}. This suggests that this portion of the correlated activity may indeed be driven by low-level visual evoked responses.  However, \revtwo{the degree of engagement, here represented by narrative coherence, appear to modulate the \emph{amplitude} of the ISC time course, since even though the scrambled stimulus was driven by the visual stimulus, it was so to a lesser extent.} 
Previous research has shown that visual evoked potentials (VEP) are modulated by spatial attention \citep{Johannes1995} and that even feature-specific attention enhances steady-state VEPs \citep{muller2006}.
We \revtwo{quantify the effect of scrambling the narrative} by comparing the sensitivity (slope) of ISC to ALD in both the normal and scrambled conditions \revone{by fitting a simple linear model} (Fig. \ref{Fig.GAINS}). For both films we found significant reductions of the ISC/ALD slope in the scrambled version ($p<0.01$; block permutation test, with block size $B = 25\ s$).

\section*{Discussion}

We have demonstrated that student \revtwo{neural reliability to} media stimuli may be quantified using EEG in a classroom setting. 
For educational technology cost and robustness are key features, hence, we aimed at establishing a realistic scenario based on low-cost consumer grade equipment, the Smartphone Brain Scanner, focusing on several potential sources that could degrade robustness.

We have provided evidence that salient aspects of the \revtwo{neural reliability previously} detected with laboratory grade equipment can be reproduced in a realistic setting. We recorded fully-synchronized EEG with nine subjects in a real classroom and found that the level of neural response reliability matched prior laboratory results.
\revone{The robustness of CorrCA and ISC is granted by the reproducibility between recording conditions, both of the ISC time-courses throughout the film clips and of the spatial topographies of the first three CorrCA components. For the \revtwo{film clip from} \textit{Bang! You're Dead} we saw that seven subjects were enough to obtain stable topographies for all three components, whereas for \textit{Sophie's Choice} and the baseline video the results were more noisy, \revtwo{suggesting} that more subjects are needed to obtain stable results.
Previous research shows that ten subjects provided for stable results in a case involving non-narrative baseline videos or films with lower ISC and IVC in a laboratory setting \citep{Dmochowski2012}.}

Mathematically, we have shown that our detection scheme, CorrCA, is robust to inter-subject variability in spatial configurations of brain networks, \revtwo{or induced by cap misalignment.}
\revone{In the calculations, we assumed two subjects \revtwo{ in a worst case scenario where the subjects' spatial projections are orthogonal.} This result conforms well with simulations \revtwo{that show} that, even for multiple subjects with randomly drawn spatial projections, CorrCA was able to find the relevant times series \citep{kamronn2015}. The simulations also showed that increasing the number of subjects decreased the signal-to-noise ratio, presumably due to the estimated common projection not being able to fit with the different projections of each subject.}

\revone{We have presented results that further \revtwo{indicate a relationship between changes in ISC and viewer engagement.} Through a basic analysis of questionnaires on scenes of high impact, we found that high ISC indeed is associated with high impact. We have also showed a relationship between neural responses to luminance fluctuations and coherence of stimulus narrative. For both the films presented, we saw a significant drop in the average IVC for subjects watching the \revtwo{film sequences in which the narrative had been temporally scrambled.}
At the same time no significant difference was found between the groups watching \revtwo{the film sequences that had not been scrambled,} which further underlines the robustness of the measure.}

It may appear surprising that there exists a significant correlation between the \emph{raw EEG signals} of various students in the classroom. However, it is well-known that eye scan patterns in a film audience follow a specific pattern after a scene change, activating the dorsal pathway  \citep{unema2005time}.
\revone{A valid assumption could therefore be that the correlation is due to synchronised artefacts from eye movements, but this has recently been shown not to affect attentional modulation of ISC \citep{ki2016attention}.
Also, it is known} that stimuli in the form of flashing images elicit VEPs, which are modulated in amplitude by the luminance  \citep{Armington1968}. When recorded with EEG, the spatial distribution of the early VEP at 100ms (P100) is similar to the scalp maps of the first correlated component (C1 in Fig. \ref{Fig.iscscalp}b)  \citep{Johannes1995,Sandmann2012}.

\revone{We investigated whether low-level visual processes could be a driving force behind the measured ISCs by correlating \revtwo{the ISC} with changes in luminance in the video stimuli, as measured by the ALD. We found that luminance fluctuations drive a significant portion of the ISC.

In all four groups of subjects \textit{Sophie's Choice} obtained lower IVC compared to \textit{Bang! You're Dead}. \revtwo{This difference could be} explained by the fact that the film clip also had a much lower ALD. Also, Fig.\ \ref{Fig.ISCscene} indicates that the passage in \textit{Bang! You're dead} with the highest and most sustained ISC (around 1:20 to 1:50) coincides \revtwo{ with the interval with the most scene changes.} This relationship could, however, also be due to more complex processes, as fast-paced cutting is a known cinematographic tool used by Hitchcock to induce suspense and thereby increase the attention of the viewer  \citep{Bordwell2002}.

The strong link between ISC and luminance fluctuations due to scene cuts have also recently been presented in a fMRI study \citep{herbec2015}. This is something that would be interesting to take into account for future studies investigating the applicability of ISC. Baseline videos could be created in ways to achieve similar ALD features as the target stimuli. The baseline video, created for this study, consisted of one continuous scene of people entering and exiting an escalator in a relaxed manner, which did not produce any significant correlation. Future studies might use a baseline video containing scene cuts of faces and body parts, to also take the effect of editing into account.

To investigate the possibility of higher level processes also being at play, we analysed the linear relationship between ISC and luminance fluctuations at a given time in the video stimulus. The scrambling operation aimed to test for a change in attentional engagement while controlling for low level features. The premise was that subjects would be less attentive to the stimulus, i.e. less "engaged", if they did not follow the narrative arch of the story. With that in mind, Fig. \ref{Fig.ISCscene} and \ref{Fig.GAINS} suggest that ISC is driven by stimulus-evoked responses that are modulated by attentional engagement with the stimulus.

\revtwo{We have demonstrated the feasibility of tracking inter-subject correlation in a classroom setting; a measure that has been related to attentional modulation \citep{ki2016attention}.
We have shown that ISC is robust to recording equipment and conditions, and we have presented evidence that the amplification of ISC in films that have a strong and coherent narrative is due to attentional modulation of visual evoked responses. Thus ISC may be used as an} indirect electrophysiological measure of engagement through an attentional top-down modulation of low-level neural processes. Recent research has shown that attentional modulation of neural responses takes place in speech perception \citep{mesgarani2012,mirkovic2015}, which lends credibility to a similar process occurring in the visual system. The evidence that such a basic and well defined mechanism could be at play further adds to the robustness of the approach in real everyday scenarios.}

\section*{Methods}
{\bf Protocol.} Four groups of subjects watched the video stimuli in different scenarios. The first group ($N=12$, \textit{Individual}) watched videos individually in an office environment on a tablet computer (Google Nexus 7 tablet, with a 7" (17.8 cm) screen) with earphones. The second group ($N=12$) saw the videos in the same manner, but the scenes of the film stimulus \revtwo{were scrambled in time resulting in the narrative being lost (\textit{Scrambled}). The objective of this condition was} to demonstrate that the similarity of responses across subjects is not simply the result of low-level stimulus features (which are identical in the \textit{Individual} and \textit{Scrambled} conditions), but instead, is modulated narrative coherence, which presumably engages viewers. Two additional groups ($N=9,\ N=9$) watched the original videos on a screen in a classroom (Figure \ref{Fig.setup_joint}, \textit{Joint 1} and \textit{Joint 2}), with sound projected through loudspeakers. An attempt was made to \revtwo{create viewing conditions for the subjects in the \textit{joint} groups, that were similar to the viewing conditions for the \textit{individual} group,} i.e., lights were dampened and the projected image produced approximately the same field-of-view (see supplementary materials).
The central question was whether the viewing condition (i.e., in a group versus individually) \revone{influences the level of ISC across subjects.}
	
{\bf Stimuli.} The first video clip was a suspenseful excerpt from the short film \textit{Bang! You're Dead} (1961) directed by Alfred Hitchcock. It was selected because it is known to elicit highly reliable brain activity across subjects in fMRI  \citep{hasson2004intersubject} as well as EEG  \citep{Dmochowski2012}.
Our second stimulus was a clip from \textit{Sophie's Choice}, \revone{directed by Alan J. Pakula (1982), which has been used earlier} to study fMRI activity in the context of emotionally salient naturalistic stimuli  \citep{Raz2012}.
A third non-narrative control video was recorded in a Danish metro station \revtwo{of several people who were being transported} quietly on an escalator.
Each video clip had a length of approximately six minutes and was shown twice to each subject. For each viewing the order of the clips was randomized, while the same random order was used the second time the clips were shown. A combined video was created for each of the six possible permutations of the order of the clips, starting with a 10 second 43 Hz tone for use in post processing synchronization, and 20 seconds black screen between each film clip. The total length of the video amounted to 39 minutes.
An additional control stimulus (\textit{Scrambled}) was created by scrambling the order of the scenes in \textit{Bang! You're Dead} and \textit{Sophie's Choice} \revtwo{in accordance with previous research}  \citep{hasson2008neurocinematics,Dmochowski2012}.
\revone{In these studies, scene segments were defined in varying temporal scales (36 s, 12 s, and 4 s) \revtwo{that consisted} of multiple camera positions, "shots". For this study we defined a scene as a single shot (i.e. the segment between two scene cuts) with the added rule \revtwo{that a scene must not exceed 250 frames ($\sim$ 10 s) to reduce subjects' ability to infer the narrative from long scenes.}
This procedure resulted in 73 scenes lasting between 0.5 and 10 seconds and corresponded to the intermediate to short time-scales employed in previous studies \citep{hasson2008neurocinematics}.}

{\bf Subjects.} A total of 42 female subjects (mean age: 22.4y, age range: 18-32y), who gave written informed consent prior to the
experiment, \revtwo{were recruited for this study.} Non-invasive experiments on healthy subjects are exempt from ethical committee processing by Danish law \citep{DenNationaleVidenskabsetiskeKomite2014}.
Among the 42 recordings, \revone{nine} were excluded due to unstable wireless communication that precluded proper synchronization of the data across subjects \revone{(five from the \textit{Individual} group, \revtwo{one from the \textit{Scrambled} group} and three from the two \textit{Joint} groups)}. \revtwo{The difference in the number of recordings in the different groups} \revone{could give unfair advantages with respect to noise when using CorrCA or calculating ISC. We therefore decided to randomly choose four subjects from the \textit{Scrambled} group and one from} \revtwo{\textit{Joint 2} group and excluded these from the analyses. This was to ensure that each group had seven fully synchronized recordings.}

{\bf Portable EEG -- Smartphone Brain Scanner.}
Research grade EEG equipment is costly, time-consuming to set up, and immobile. However, recently consumer grade EEG equipment \revtwo{that is more affordable and has increased comfort has appeared.} Here we use the modified 14 channel system, 'Emocap', based on the EEG Emotiv EPOC headset.  For details and validation, see  \citep{stopczynski2014smartphone,stopczynski2014a}. \revtwo{In this study it was implemented} on Asus Nexus 7 tablets. An electrical trigger and associated sound was used to synchronize EEG and video signals in the individual viewing condition, while a split audio signal (simultaneously feeding into microphone and EEG amplifiers) was used to synchronize the nine subjects EEG recordings and the video in the joint viewing condition
\revone{(see supplementary materials for further information on synchronisation).}
The resulting timing uncertainty was measured to be less than 16 ms.
\revone{The EEG was recorded at 128 Hz and subsequently bandpass filtered digitally}
using a linear phase windowed sinc FIR filter between 0.5 and 45 Hz and shifted to adjust for group delay. Eye artefacts were reduced with  a conservative pre-processing procedure using independent component analysis (ICA), removing up to 3 of the 14 available components (Corrmap plug-in for EEGLAB  \citep{Delorme2004,Viola2009}).

{\bf Correlated component analysis to measure ISC and IVC.}
CorrCA was presented in Dmochowski et al. 2012, as a constrained version of Canonical Correlation Analysis (CCA).
\revone{CorrCA seeks to find sets of weights that maximises the correlation between the neural activity of subjects experiencing the same stimuli. For each neural component, CorrCA finds one shared
set of weights for all subjects in the group.}

Given two multivariate spatio-temporal time series \revtwo{(termed “view” in CorrCA)}, $\{\textbf{X}_{1},\textbf{X}_{2}\} \in \mathbb{R}^{D\times N}$, with $D$ being the number of measured features (EEG channels) in the two views and $N$ the number of time samples, CCA estimates weights, $\{\textbf{w}_{1},\textbf{w}_{2}\}$, which maximize the correlation between the components, $\textbf{y}_1 = \textbf{X}_1^\intercal\textbf{w}_1$ and $\textbf{y}_2 = \textbf{X}_2^\intercal\textbf{w}_2$. The weights are calculated using two eigenvalue equations, with the constraint that the components \revone{belonging to each multivariate time series} are uncorrelated  \citep{Hardoon2004}. CorrCA is relevant for the case where the views are homogeneous, e.g., using the same EEG channel positions, and imposes the additional constraint of shared weights $\textbf{w} = \textbf{w}_1 = \textbf{w}_2$. This assumption can potentially increase sensitivity involving fewer parameters. In CorrCA the weights are thus estimated through a single eigenvalue problem;

\vspace{-10pt}
\begin{align}
\left(\textbf{R}_{11} + \textbf{R}_{22}\right)&^{-1} \left(\textbf{R}_{12} + \textbf{R}_{21}\right)\textbf{w} = \rho \textbf{w},\label{eq.correlated component analysis}
\end{align}
where, $\textbf{R}_{ij}=\frac{1}{N}\textbf{X}_i\textbf{X}_j^\intercal$, is the sample covariance matrix  \citep{Dmochowski2012}. \revone{To illustrate the spatial distribution of the underlying physiological activity of the components, we use the estimated forward models ("patterns") as discussed in  \citep{Parra2005,haufe2014}.}

{\bf Average luminance difference (ALD).} Video clips were converted to grey scale (0-255) by averaging over the three colour channels. We then calculated the squared difference in pixel intensity from one frame to the next and took the average across pixels. These signals were non-linearly re-sampled at 1Hz by selecting the maximum ALD for each 1 $s$ interval \revone{to emphasise the large differences during changes in camera position} (see figure \revone{S2} in supplementary materials for an comparison between frame-to-frame and smoothed difference). These values were then smoothed in time by convolving with a Gaussian kernel with a "variance" parameter of 2.5 $s^2$. This down sampling and smoothing was aimed at matching the temporal resolution of the ALD to that of the time-resolved ISC computation (5 $s$ sliding window with 1 $s$ intervals). 

{\bf Statistical testing.}
In order to evaluate the statistical relevance of the correlations, we employed a simple permutation test ($P=5000$ permutations) \citep{Dmochowski2012}. \revtwo{To test the robustness of the obtained weights for the spatial projections, we calculated the average correlation of all possible pairings of the four conditions groups for a given component. Again, we employed a permutation test ($P=5000$ permutations) to evaluate statistical relevance by randomly permuting the channel order for each group and recalculating the average correlation.} When testing differences in average IVC between conditions, \revtwo{we used a block permutation test (block size $B = 25\ s$, $P=5000$ permutations) to account for temporal dependencies.}

\bibliographystyle{apa}
\bibliography{biblio}

\section*{Acknowledgements}
We thank Ivana Konvalinka, Arek Stopczynski and the DTU Smartphone Brain Scanner team for their assistance and helpful discussions. This work was supported by the Lundbeck Foundation through the {\it Center for Integrated Molecular Brain Imaging} and by Innovation Foundation Denmark through {\it Neurotechnology for 24/7 brain state monitoring}.

\section*{Author contributions statement}
ATP, SK, JD, LP and LKH designed research; SK, ATP and LKH performed research;
  ATP, SK, JD, LP, and LKH contributed analytical tools;  ATP, SK, LKH  analysed data;  ATP, SK, JD, LP, and LKH wrote the paper.

\section*{Additional information}

\textbf{Competing financial interests}
 The authors declare that the research was conducted in the absence of any commercial or financial
 relationships that could be construed as a potential conflict of interest.

\end{document}